\documentclass[aps,prd,twocolumn,superscriptaddress,showpacs]{revtex4}
\usepackage{epsfig,epsf}
\usepackage{amsmath}
\usepackage{amsthm}
\usepackage{amsfonts}
\usepackage{amssymb}
\usepackage{dsfont}
\usepackage{multirow}
\usepackage{appendix}
\usepackage{slashed}
\usepackage[active]{srcltx}
\usepackage{psfrag}
\usepackage{subfigure}
\newcommand{\be}{\begin{equation}}
\newcommand{\ee}{\end{equation}}
\newcommand{\ba}{\begin{eqnarray}}
\newcommand{\ea}{\end{eqnarray}}
\newcommand{\baa}{\begin{eqnarray*}}
\newcommand{\btab}{\begin{tabular}}
\newcommand{\etab}{\end{tabular}}
\newcommand{\eaa}{\end{eqnarray*}}

\def\inbar{\,\vrule height1.5ex width.4pt depth0pt}
\def\IC{\relax\hbox{$\inbar\kern-.3em{\rm C}$}}
\def\IZ{\relax{\hbox{\cmss Z\kern-.4em Z}}}
\def\IR{{\hbox{{\rm I}\kern-.2em\hbox{\rm R}}}}

\def\IP{{\hbox{{\rm I}\kern-.2em\hbox{\rm P}}}}
\def\II{\hbox{{1}\kern-.25em\hbox{l}}}

\begin{document}

\title{Impact of finite density on spectroscopic parameters of decuplet baryons }
\date{\today}
\author{K. Azizi$^1$, N. Er$^2$, H. Sundu$^3$  \\
\textit{$^1$ Department of Physics, Do\v{g}u\c{s} University, Ac{\i}badem-Kad{\i}k\"oy, 34722 Istanbul, Turkey}\\
\textit{$^2$ Department of Physics, Abant \.{I}zzet Baysal University,
G\"olk\"oy Kamp\"us\"u, 14980 Bolu, Turkey}\\
\textit{$^3$ Department of Physics, Kocaeli University,  41380 \.{I}zmit, Turkey}}

\begin{abstract}
The decuplet baryons, $\Delta$, $\Sigma^{*}$,  $\Xi^{*}$ and $\Omega^{-}$, are studied in nuclear matter by using the in-medium QCD sum rules. 
By fixing the three momentum of the particles under consideration at the rest frame of the medium, the negative energy contributions are removed. It is obtained that the 
parameters of the  $\Delta$ baryon are more affected by the medium against the $\Omega^{-}$ state, containing three strange quarks, whose mass and residue do not affected by the medium, considerably. 
We also find the vector and scalar self energies of these baryons in nuclear matter. By the recent progresses at $\bar{P}$ANDA experiment at FAIR
 and NICA facility it may be possible to study the in-medium properties of such states even the multi-strange $\Xi^{*}$ and $\Omega^{-}$ systems in near future.
\end{abstract}

% nuclear matter, baryons, Protons and neutrons, Hyperons, sum rules
\maketitle

%%%%%%%%%%%%%%%%%%%%%%%%%%%%%%%%%%%%%%%%%%%%%%%%%%%%%%%%%%%%%%%%%%%%%%%%%%%%

%\new page

%%%%%%%%%%%%%%%%%%%%%%%%%%%%%%%%%%%%%%%%%%%%%%%%%%%%%%%%%%%%%%%%%%%%%%%%%%%%%%%%%%%%%%%%%%%%%%%%%%%
%

\section{Introduction}
The investigations  of the  properties of hadrons under extreme conditions have been in the focus of much attention for many years. Such investigations are very important in the study of the internal 
 structure of the dense astrophysical objects like neutron stars. The formation of neutron stars is influenced 
by all four known fundamental interactions. Hence, understanding of their nature can help us  in the course of unification of all fundamental forces within a common theoretical framework, which is 
one of the biggest challenges for physics.
The recent observation of massive neutron stars with roughly  twice the solar mass \cite{Demorest,Antoniadis} has stimulated the focuses on the equation of state of the dense nuclear matter
(see for instance \cite{Bastian:2015avq,Fraga:2013qra,Kim:2016yem,Drews:2014spa}). 
However, the expected appearance of hyperons at about two times nuclear density, called  ``hyperon puzzle'' remains an unresolved mystery in neutron stars 
(concerning the appearance of hyperons in neutron stars see for example \cite{Weissenborn:2011ut, Weissenborn:2011kb}).
It has also found  that $\Delta$ isobars appear at a density of the order of $2\div3$ times nuclear matter saturation density, and a `` $\Delta$ puzzle'' exists, similar to the ''hyperon puzzle''
if the potential of the  $\Delta$ in nuclear matter is close to the one indicated by the experimental data \cite{Drago:2014oja}. More theoretical and experimental  investigations on the properties
of strange and non-strange light baryons in dense medium are needed to solve such puzzles.

From the experimental side, the bound nuclear systems with one, two or three  units of strangeness are  poorly known compared to that of the non-strange states like nucleons. 
The  large production probability of various hyperon-antihyperon pairs in antiproton collisions will provide opportunities for  series of new studies on the behavior of the  systems  containing 
two or even more units of strangeness at the $\bar{P}$ANDA experiment at FAIR. By the progresses made, it will be possible to study 
the in-medium properties of the doubly strange $\Lambda\Lambda$-hypernuclei as well as  the multi-strange $\Xi^-$, $\bar{\Xi}^+$ and $\Omega^{-}$ systems in near future \cite{Singh:2016hoh}.

From the theoretical side,  the effects of nuclear medium on 
the physical parameters of the nucleon  have been widely investigated in the literature (see for instance \cite{Cohen:1994wm,Drukarev88,Hatsuda91,Adami91,Azizi:2014yea} and references therein). 
But, we have only  a few studies dedicated to the in-medium properties of hyperons and decuplet baryons in the literature 
(for instance see \cite{Azizi:2015ica,Jin94,Jin95,Savage96,Miyatsu2009,Beane2012,Ouellette:1997ip}). In the present study, we investigate the impact of nuclear matter on some spectroscopic parameters of the 
$\Delta$, $\Sigma^{*}$,  $\Xi^{*}$ and $\Omega^{-}$ decuplet baryons. In particular, we calculate the mass and  residue as well as the scalar and vector self energies of these baryons using the 
well established in-medium QCD sum rule approach. We compare the in-medium results  with those obtained at $\rho=0$ or vacuum and find the corresponding shifts. To remove the contributions of the negative energy 
particles, we work at the rest frame of the nuclear matter and fix the three momentum of the particles under consideration.

\section{ $\Delta$, $\Sigma^{*}$,  $\Xi^{*}$ and $\Omega^{-}$  baryons in nuclear matter }

In this section we aim to construct sum rules for the mass, residue and vector self energy of the decuplet baryons and numerically analyze the obtained results. To this end and in accordance with the
 general philosophy of the QCD sum rule approach,we start with a correlation function as the building block of the method:
\begin{equation}\label{correlationfunc}
\Pi_{\mu\nu}(p)=i\int{d^4 xe^{ip\cdot
x}\langle\psi_0|T[\eta_{\mu,D}
(x)\bar{\eta}_{\nu,D}(0)]|\psi_0\rangle},
\end{equation}
where $p$ is the four momentum of the decuplet ($D$) baryon, $|\psi_0\rangle$ is the  ground state of the nuclear matter and $\eta_{\mu,D}$ is the interpolating current of the $D$ baryon. 
 The  general form of the interpolating current for decuplet baryons in a compact form reads ,
\begin{eqnarray}\label{}
\eta_{\mu,D}&=&A_D \epsilon^{abc} \Big\{(q_{1}^{aT}C\gamma_\mu q_{2}^{b})q_{3}^{c} + (q_{2}^{aT}C\gamma_\mu q_{3}^{b})q_{1}^{c} \nonumber \\
&+&  (q_{3}^{aT}C\gamma_\mu q_{1}^{b})q_{2}^{c} \Big\},
\end{eqnarray}
where $a, b, c$ are color indices, $C$ is the charge conjugation operator and $A_D$ is the normalization constant. The quark content and  value of $A_D$ for different members are given 
in table I \cite{Chung:1981cc}.
\begin{table}
\renewcommand{\arraystretch}{1.3}
\addtolength{\arraycolsep}{-0.5pt}
\small
\begin{tabular}{|c|c|c|c|c|}\hline \hline  &$A_D$ & $q_{1}$ & $q_{2}$ & $q_{3}$  \\\hline
$\Sigma^{*}$ & $\sqrt{2/3}$ & u & d & s \\
$\Delta^0$ & $\sqrt{1/3}$ & d & d & u \\
$\Xi^{*}$ & $\sqrt{1/3}$ & s & s & u\\
$\Omega^{-}$ & $1/3$ & s & s & s\\ \hline \hline \end{tabular}
\caption{The value of the normalization constant $A_D$ and the quark flavors $q_1$, $q_2$, $q_3$ for the decuplet baryons.}
\renewcommand{\arraystretch}{1}
\addtolength{\arraycolsep}{-1.0pt}
\end{table}
We will calculate the aforementioned correlation function in two representations: hadronic and OPE (operator product expansion). By equating these two representations, one
can get the QCD sum rules for the aimed physical quantities.

\subsection{Hadronic Representation}
The correlation function in the hadronic side is obtained by inserting a complete set of baryonic state with the same quantum numbers as the interpolating current. After performing the integral over four-$x$, we get
\begin{eqnarray}\label{phepi}
\Pi_{\mu\nu}^{Had}(p)&=&-\frac{{\langle}\psi_0|\eta_{\mu,D}(0)|D(p^*,s){\rangle}
{\langle}D(p^*, s)|\bar{\eta}_{\nu,D}(0)|\psi_0{\rangle}}{p^{*2}-m_{D}^{*2}} \notag \\
&+&...,
\end{eqnarray}
where $|D(p^*,s){\rangle}$ is the decuplet baryon state with spin $s$ and in-medium four momentum $p^*$, $m_D^*$ is the modified mass of the decuplet baryon in medium 
and $...$ indicates the contributions of the higher states and  continuum. The matrix elements in Eq. (\ref{phepi}) can be represented as
\begin{eqnarray}\label{spinor}
{\langle}\psi_0|\eta_{\mu,D}(0)|D(p^*,s){\rangle} &=& \lambda_{D}^{*} u_{\mu} (p^*,s), \nonumber \\
{\langle}D(p^*, s)|\bar{\eta}_{\nu,D}(0)|\psi_0{\rangle} &=& \bar{\lambda}_{D}^{*} \bar{u}_{\nu} (p^*,s),
\end{eqnarray}
where $u_{\mu} (p^*,s)$ is the in-medium Rarita-Schwinger spinor and $\lambda_{D}^{*} $ is the modified residue or the coupling strength of the decuplet baryon to nuclear medium. 
 Inserting Eq. (\ref{spinor}) into Eq. (\ref{phepi}) and summing over the spins of the $D$ baryon one can, in principle, find the hadronic side of the correlation function. 
Before that, it should be remarked that the current $\eta_{\mu,D}$ couples to both the spin-$1/2$ octet states and the spin-$3/2$ decuplet states. 
In order to get only the contributions of the decuplet baryons, the contributions of the unwanted  spin-$1/2$ states must be removed from the correlation function. 
For this aim, we come next with  the following procedure. The matrix element of $\eta_{\mu,D}$ between the spin-$1/2$ and in-medium states can be decomposed as
\begin{equation}\label{}
{\langle}\psi_0|\eta_{\mu,D}(0)|\frac{1}{2} (p^*){\rangle} = \Big (C_1 p^*_{\mu} + C_2 \gamma_{\mu} \Big ) u (p^*),
\end{equation}
where $C_1$ and $C_2$ are constants and $u (p^*)$ is the in-medium Dirac spinor of momentum $p^*$. 
By multiply both sides of the above equation with $\gamma^{\mu}$ and   using the condition $\eta_{\mu,D} \gamma^{\mu}=0$, we immediately find the constant $C_1$ in terms of $C_2$. Hence, 
\begin{equation}\label{}
{\langle}\psi_0|\eta_{\mu,D}(0)|\frac{1}{2} (p^*){\rangle} = C_2 \Big (-\frac{4} {m_{1/2}^{*}}p^*_{\mu} +  \gamma_{\mu} \Big ) u (p^*),
\end{equation}
where $m_{1/2}^{*}$ is the modified mass of the spin-$1/2$ baryons. It can be easily seen that the unwanted contributions of the spin-$1/2$ states are proportional to
  $p^*_{\mu} $ and $ \gamma_{\mu}$. By ordering the Dirac matrices as $\gamma_{\mu}\!\not\!{p^*}\gamma_{\nu}$ and setting to zero the terms with $\gamma_{\mu}$ in the beginning and $\gamma_{\nu}$ at the
 end and those  proportional to $p^*_{\mu}$ and $p^*_{\nu}$, the contributions from the unwanted spin-$1/2$ states can be easily eliminated.

Now, we insert  Eq. (\ref{spinor}) into   Eq. (\ref{phepi}) and use the  summation over spins of the Rarita-Schwinger spinor as
\begin{eqnarray}\label{Rarita}
\sum_s  u_{\mu} (p^*,s)  \bar{u}_{\nu} (p^*,s) &= &-(\!\not\!{p^*} + m^{*}_{D})\Big[g_{\mu\nu} -\frac{1}{3} \gamma_{\mu} \gamma_{\nu} \nonumber \\
&-& \frac{2p^*_{\mu}p^*_{\nu}}{3m^{*2}_{D}} +\frac{p^*_{\mu}\gamma_{\nu}-p^*_{\nu}\gamma_{\mu}}{3m^{*}_{D}} \Big],
\end{eqnarray}
as a result of which we get
\begin{eqnarray}\label{}
\Pi_{\mu\nu}^{Had}(p)&=&\frac{\lambda_{D}^{*} \bar{\lambda}_{D}^{*}(\!\not\!{p^*} + m^{*}_{D})}{p^{*2}-m_{D}^{*2}} \Big[g_{\mu\nu} -\frac{1}{3} \gamma_{\mu} \gamma_{\nu} \nonumber \\
&-& \frac{2p^*_{\mu}p^*_{\nu}}{3m^{*2}_{D}} +\frac{p^*_{\mu}\gamma_{\nu}-p^*_{\nu}\gamma_{\mu}}{3m^{*}_{D}} \Big]+ ....
\end{eqnarray}

To proceed, we would like to mention that  the in-medium momentum and the modified mass can be written in terms of the self energies $\Sigma_{\mu, \nu}$ and $\Sigma^S$  as  $p_{\mu}^*=p_{\mu}-\Sigma_{\mu, v}$  and $m^{*}_D=m_D+\Sigma^S$, where  $\Sigma^S$ is the scalar self energy. The  self-energy   $\Sigma_{\mu, v}$ can also be written in a general form as
\begin{equation}\label{}
\Sigma_{\mu,v}=\Sigma_{v} u_{\mu} + \Sigma'_{v}p_{\mu}
\end{equation}
where $\Sigma_{v} $ is called the vector self energy  and $u_{\mu}$ is the four velocity of the nuclear medium.  In the mean-field approximation, the scalar and vector self energies are obtained to be real and independent of momentum and  the $ \Sigma'_{\nu}$ is taken to be identically zero \cite{Cohen:1994wm,Thomas:2007gx}. In this context, particles of any three-momentum  appear as stable quasi-particles  with self energies that are roughly linear in the density up to nuclear matter density \cite{Cohen:1994wm,Serot:1984ey}.
We perform the calculations in the rest frame of the nuclear medium, i.e. $u_{\mu}=(1,0)$ and at fixed
three-momentum of D baryon, $|\vec{p}|$. We get
 \begin{eqnarray}\label{}
\Pi_{\mu\nu}^{Had}(p_0, \vec{p})&=& \lambda_{D}^{*} \bar{\lambda}_{D}^{*}\frac{(\!\not\!{p}-\Sigma_{v}\!\not\!{u}+m^{*}_D)}{p^2+\Sigma_{v}^2 - 2p_0\Sigma_{v}-m^{*2}_D}\Big[  g_{\mu\nu} \nonumber \\
&&-\frac{1}{3}\gamma_{\mu}\gamma_{\nu} -\frac{2}{3 m^{*2}_D}\Big(p_{\mu}p_{\nu}-\Sigma_{v}p_{\mu}u_{\nu} \nonumber \\
&&-\Sigma_{v}u_{\mu}p_{\nu}+\Sigma^2_{v}u_{\mu}u_{\nu}\Big) +\frac{1}{3 m^{*}_D}\Big(p_{\mu}\gamma_{\nu} \nonumber \\
&&-\Sigma_{v}u_{\mu}\gamma_{\nu}-p_{\nu}\gamma_{\mu}+\Sigma_{v}u_{\nu}\gamma_{\mu}\Big)\Big] \nonumber \\&&+ ...,
\end{eqnarray}
where $p_0=p\cdot u$ is the energy of the quasi-particle. After  ordering of the Dirac matrices 
  and eliminating the unwanted spin-1/2 contributions, we get
\begin{eqnarray}\label{}
\Pi_{\mu\nu}^{Had}(p_0, \vec{p})&=&
\frac{\lambda_{D}^{*} \bar{\lambda}_{D}^{*}}{(p_0-E_p)(p_0-\bar{E}_p)}\Big[ m^{*}_{D} g_{\mu\nu}+g_{\mu\nu}\!\not\!{p} \nonumber \\
&&-\Sigma_{v}g_{\mu\nu}\!\not\!{u}  \Big]  + ...,
\end{eqnarray}
where ,
$E_p=\Sigma_{v}+\sqrt{|\vec{p}|^2+ m^{*2}_{D}}$ and
$\bar{E}_p=\Sigma_{v}-\sqrt{|\vec{p}|^2+ m^{*2}_{D}}$ are the positions of the positive- and negative energy poles, respectively. One can write the above equation as an
integral representation in terms of the spectral
density,
\begin{equation}\label{}
\Pi_{\mu\nu}^{Had}(p_0, \vec{p})=\frac{1}{2\pi i} \int_{-\infty}^{\infty} d\omega \frac{\Delta \rho_{\mu\nu}^{Had}(p_0, \vec{p})}{\omega-p_0}
\end{equation}
where the spectral density $ \Delta \rho_{\mu\nu}^{Had}(p_0, \vec{p}) $, defining 
by 
\begin{eqnarray}\label{shwartz}
 &&\Delta \rho_{\mu\nu}^{Had}(p_0, \vec{p})\nonumber \\&&=Lim_{\epsilon\rightarrow0^+}[\Pi_{\mu\nu}^{Had}(\omega+i\epsilon, \vec{p})-\Pi_{\mu\nu}^{Had}(\omega-i\epsilon, \vec{p})], \nonumber \\
\end{eqnarray}
is given as
\begin{eqnarray}\label{}
\Delta \rho_{\mu\nu}^{Had}(p_0, \vec{p})&=&-\frac{1}{2\sqrt{m_{D}^{*2}+|\vec{p}|^2}} \lambda_{D}^{*} \bar{\lambda}_{D}^{*}\Big[ m^{*}_{D} g_{\mu\nu}+g_{\mu\nu}\!\not\!{p} \nonumber \\
&&-\Sigma_{v}g_{\mu\nu}\!\not\!{u} \Big] \Big[ \delta(\omega - E_p) - \delta(\omega- \bar{E}_p) \Big] . \nonumber \\
\end{eqnarray}
The next step is to exclude the negative-energy pole contribution by multiplying the correlation function
  with the
weight function $(\omega-\bar{E}_p)e^{\frac{-\omega^2}{M^2}}$ and performing the integral over $ \omega $ from $ -\omega_0 $ to $ \omega_0 $, i.e.
\begin{equation}\label{integ}
\Pi_{\mu\nu}^{Had}(p_0, \vec{p})= \int_{-\omega_0}^{\omega_0}
d\omega \Delta \rho_{\mu\nu}^{Had}(\omega,
\vec{p})(\omega-\bar{E}_p)e^{-\frac{\omega^2}{M^2}},
\end{equation}
where $\omega_0$ is the threshold parameter and $M^2$ is the Borel mass parameter which shall be fixed later. After performing the integral in Eq. (\ref{integ}),  the hadronic side of the correlation function takes its final form in terms of the corresponding structures,
\begin{eqnarray}
\Pi_{\mu\nu}^{Had}(p_0, \vec{p})&=& \lambda_{D}^{*2}  e^{-E^2_p/M^2}
\Big[ m^{*}_{D} g_{\mu\nu}+g_{\mu\nu}\!\not\!{p} -\Sigma_{v}g_{\mu\nu}\!\not\!{u}  \Big]. \nonumber \\
\end{eqnarray}

%We will choose the structures $g_{\mu\nu}$,
%$\!\not\!{p}g_{\mu\nu}$ and $\!\not\!{u}g_{\mu\nu}$ to find the
%mass residue and self-energies for decuplet baryons.
%%%%%%%%%%%%%%%%%%%%%%%%%%%%%%%%%%%%%%%%%%%%%%%%%%%%%%%%%%%%%%%%%%%%%%%%%%%%%%%%%%%%%%%%%%%%%%%%%%%%

\subsection{OPE Representation}

The OPE side of the correlation function is calculated at the large space-like region  $ p^2\ll0 $ 
in terms of QCD degrees of freedom. One can write the OPE side of the correlation function, in terms
of the involved structures, as
\begin{eqnarray}\label{eq1}
\Pi_{\mu\nu}^{OPE}(p_0,\vec{p})&=&\Pi_{1}(p_0,\vec{p})g_{\mu\nu}+\Pi_{2}(p_0,\vec{p})\!\not\!{p}
g_{\mu\nu}
\nonumber \\
&+&\Pi_{3}(p_0,\vec{p})\!\not\!{u}g_{\mu\nu},\nonumber \\
\end{eqnarray}
where the $\Pi_{i}(p_0,\vec{p})$ functions, with $i=1,2$ or $3$, can be
written in terms of the spectral densities $\Delta \rho_i(p_0,\vec{p})$ in OPE side
as
\begin{eqnarray}\label{eq1}
\Pi_{i}(p_0,\vec{p})&=&\frac{1}{2\pi i}\int^{\infty}_{-\infty} dw
\frac{\Delta \rho_i(p_0,\vec{p})}{w-p_0},
\end{eqnarray}
where $\Delta \rho_i(p_0,\vec{p})$ are the imaginary parts of
$\Pi_{i}(p_0,\vec{p})$ functions obtaining from the OPE version of Eq. (\ref{shwartz}). The main aim, in the present subsection,  is to find the
$\Delta\rho_i(p_0,\vec{p})$ spectral densities, by using of which
we can find the $\Pi_{i}(p_0,\vec{p})$ functions in OPE side. To
proceed, we start with the correlation function in Eq.
(\ref{correlationfunc}). By substituting the
explicit form of the interpolating current for the decuplet baryons under consideration into the correlation
function in Eq. (\ref{correlationfunc}) and after contracting out all
the quark pairs using the Wick's theorem,  we get 
\begin{widetext}
\begin{eqnarray}\label{corre1}
\Pi_{\mu\nu}^{OPE,\Delta}(p) &=& \frac{i}{3}\epsilon_{abc}\epsilon_{a'b'c'}\int d^4 x e^{ipx} \Bigg\langle \Bigg\{2S^{ca'}_{d}(x)\gamma_{\nu}S'^{ab'}_{d}(x)\gamma_{\mu}S^{bc'}_{u}(x)-2S^{cb'}_{d}(x)\gamma_{\nu}S'^{aa'}_{d}(x)\gamma_{\mu}S^{bc'}_{u}(x)\nonumber\\
&+&4S^{cb'}_{d}(x)\gamma_{\nu}S'^{ba'}_{u}(x)\gamma_{\mu}S^{ac'}_{d}(x)+2S^{ca'}_{u}(x)\gamma_{\nu}S'^{ab'}_{d}(x)\gamma_{\mu}S^{bc'}_{d}(x)\nonumber\\
&-&2S^{ca'}_{u}(x)\gamma_{\nu}S'^{bb'}_{d}(x)\gamma_{\mu}S^{ac'}_{d}(x)-S^{cc'}_{u}(x)Tr\Bigg[S^{ba'}_{d}(x)\gamma_{\nu}S'^{ab'}_{d}(x)\gamma_{\mu}\Bigg]\nonumber\\
&+&S^{cc'}_{u}(x)Tr\Bigg[S^{bb'}_{d}(x)\gamma_{\nu}S'^{aa'}_{d}(x)\gamma_{\mu}\Bigg]-4S^{cc'}_{d}(x)Tr\Bigg[S^{ba'}_{u}(x)\gamma_{\nu}S'^{ab'}_{d}(x)\gamma_{\mu}\Bigg]\Bigg\}
\Bigg\rangle,
\end{eqnarray}
\begin{eqnarray}\label{corre2}
\Pi_{\mu\nu}^{OPE,\Sigma^{*}}(p) &=& -\frac{2i}{3}\epsilon_{abc}\epsilon_{a'b'c'}\int d^4 x e^{ipx}\Bigg\langle \Bigg\{S^{ca'}_{d}(x)\gamma_{\nu}S'^{bb'}_{u}(x)\gamma_{\mu}S^{ac'}_{s}(x)\nonumber\\
&+&S^{cb'}_{d}(x)\gamma_{\nu}S'^{aa'}_{s}(x)\gamma_{\mu}S^{bc'}_{u}(x)+S^{ca'}_{s}(x)\gamma_{\nu}S'^{bb'}_{d}(x)\gamma_{\mu}S^{ac'}_{u}(x)\nonumber\\
&+&S^{cb'}_{s}(x)\gamma_{\nu}S'^{aa'}_{u}(x)\gamma_{\mu}S^{bc'}_{d}(x)+S^{ca'}_{u}(x)\gamma_{\nu}S'^{bb'}_{s}(x)\gamma_{\mu}S^{ac'}_{d}(x)\nonumber\\
&+&S^{cb'}_{u}(x)\gamma_{\nu}S'^{aa'}_{d}(x)\gamma_{\mu}S^{bc'}_{s}(x)+S^{cc'}_{s}(x)Tr\Bigg[S^{ba'}_{d}(x)\gamma_{\nu}S'^{ab'}_{u}(x)\gamma_{\mu}\Bigg]\nonumber\\
&+&S^{cc'}_{u}(x)Tr\Bigg[S^{ba'}_{s}(x)\gamma_{\nu}S'^{ab'}_{d}(x)\gamma_{\mu}\Bigg]+S^{cc'}_{d}(x)Tr\Bigg[S^{ba'}_{u}(x)\gamma_{\nu}S'^{ab'}_{s}(x)\gamma_{\mu}\Bigg]\Bigg\}
\Bigg\rangle,
\end{eqnarray}
\begin{eqnarray}\label{corre3}
\Pi_{\mu\nu}^{OPE,\Xi^{*}}(p) &=& \frac{i}{3}\epsilon_{abc}\epsilon_{a'b'c'}\int d^4 x e^{ipx}\Bigg\langle \Bigg\{2S^{ca'}_{s}(x)\gamma_{\nu}S'^{ab'}_{s}(x)\gamma_{\mu}S^{bc'}_{u}(x)\nonumber\\
&-&2S^{cb'}_{s}(x)\gamma_{\nu}S'^{aa'}_{s}(x)\gamma_{\mu}S^{bc'}_{u}(x)+4S^{cb'}_{s}(x)\gamma_{\nu}S'^{ba'}_{u}(x)\gamma_{\mu}S^{ac'}_{s}(x)\nonumber\\
&+&2S^{ca'}_{u}(x)\gamma_{\nu}S'^{ab'}_{s}(x)\gamma_{\mu}S^{bc'}_{s}(x)-2S^{ca'}_{u}(x)\gamma_{\nu}S'^{bb'}_{s}(x)\gamma_{\mu}S^{ac'}_{s}(x)\nonumber\\
&-&S^{cc'}_{u}(x)Tr\Bigg[S^{ba'}_{s}(x)\gamma_{\nu}S'^{ab'}_{s}(x)\gamma_{\mu}\Bigg]+S^{cc'}_{u}(x)Tr\Bigg[S^{bb'}_{s}(x)\gamma_{\nu}S'^{aa'}_{s}(x)\gamma_{\mu}\Bigg]\nonumber\\
&-&4S^{cc'}_{s}(x)Tr\Bigg[S^{ba'}_{u}(x)\gamma_{\nu}S'^{ab'}_{s}(x)\gamma_{\mu}\Bigg]\Bigg\}
\Bigg\rangle,
\end{eqnarray}
and
\begin{eqnarray}\label{corre4}
\Pi_{\mu\nu}^{OPE,\Omega^{-}}(p) &=& \epsilon_{abc}\epsilon_{a'b'c'}\int d^4 x e^{ipx}\Bigg\langle \Bigg\{S^{ca'}_{s}(x)\gamma_{\nu}S'^{ab'}_{s}(x)\gamma_{\mu}S^{bc'}_{s}(x)\nonumber\\
&-&S^{ca'}_{s}(x)\gamma_{\nu}S'^{bb'}_{s}(x)\gamma_{\mu}S^{ac'}_{s}(x)-S^{cb'}_{s}(x)\gamma_{\nu}S'^{aa'}_{s}(x)\gamma_{\mu}S^{bc'}_{s}(x)\nonumber\\
&+&S^{cb'}_{s}(x)\gamma_{\nu}S'^{ba'}_{s}(x)\gamma_{\mu}S^{ac'}_{s}(x)-S^{cc'}_{s}(x)Tr\Bigg[S^{ba'}_{s}(x)\gamma_{\nu}S'^{ab'}_{s}(x)\gamma_{\mu}\Bigg]\nonumber\\
&+&S^{cc'}_{s}(x)Tr\Bigg[S^{bb'}_{s}(x)\gamma_{\nu}S'^{aa'}_{s}(x)\gamma_{\mu}\Bigg]\Bigg\}
\Bigg\rangle,
\end{eqnarray}

\end{widetext}
where $S'=CS^TC$. Here, $S_{u,d, s}$ denotes the light quark propagator and it is given  at the
nuclear medium in the fixed-point gauge as
\cite{Cohen:1994wm}
\begin{eqnarray}\label{propagator}
 S_{q}^{ab}(x)&\equiv& \langle\psi_0|T[q^a
(x)\bar{q}^b(0)]|\psi_0\rangle_{\rho_N} \nonumber \\
&=& \frac{i}{2\pi^2}\delta^{ab}\frac{1}{(x^2)^2}\not\!x
-\frac{m_q }{ 4\pi^2} \delta^ { ab } \frac { 1}{x^2}+\chi^a_q(x)\bar{\chi}^b_q(0) \nonumber \\
&-& \frac{ig_s}{32\pi^2}F_{\mu\nu}^A(0)t^{ab,A
}\frac{1}{x^2}[\not\!x\sigma^{\mu\nu}+\sigma^{\mu\nu}\not\!x]+...,\nonumber \\
\end{eqnarray}
where $\rho_N$ is the nuclear matter density, $m_q$ is the light
quark mass, $\chi^a_q$ and $\bar{\chi}^b_q$ are the Grassmann
background quark fields and $F_{\mu\nu}^A$ are classical
background gluon fields. After inserting Eq. (\ref{propagator}) in
Eq. (\ref{corre1}) - Eq. (\ref{corre4}), we  obtain the products
of the Grassmann background quark fields and classical background
gluon fields which correspond to the ground-state matrix elements
of the corresponding quark and gluon operators \cite{Cohen:1994wm}
\begin{eqnarray}\label{}
\chi_{a\alpha}^{q}(x)\bar{\chi}_{b\beta}^{q}(0)&=&\langle
q_{a\alpha}(x)\bar{q}_{ b\beta}(0)\rangle_{\rho_N},
 \nonumber \\
F_{\kappa\lambda}^{A}F_{\mu\nu}^{B}&=&\langle
G_{\kappa\lambda}^{A}G_{\mu\nu}^{B}\rangle_{\rho_N}, \nonumber \\
\chi_{a\alpha}^{q}\bar{\chi}_{b\beta}^{q}F_{\mu\nu}^{A}&=&\langle
q_{a\alpha}\bar{q}_{ b\beta}G_{\mu\nu}^{A}\rangle_{\rho_N},
 \nonumber \\
 \mbox{and }
 \nonumber \\
\chi_{a\alpha}^{q}\bar{\chi}_{b\beta}^{q}\chi_{c\gamma}^{q}\bar
{\chi}_{d\delta}^{q}&=&\langle q_{a\alpha}\bar{q}_{b\beta}
q_{c\gamma}\bar{q}_{d\delta}\rangle_{\rho_N}.
\end{eqnarray}

Now, we need to define the quark, gluon and mixed
condensates in nuclear matter. The matrix element $\langle
q_{a\alpha}(x)\bar{q}_{b\beta}(0)\rangle_{\rho_N}$ is parameterized 
 as \cite{Cohen:1994wm}
\begin{eqnarray} \label{ }
\langle
q_{a\alpha}(x)\bar{q}_{b\beta}(0)\rangle_{\rho_N}&=&-\frac{\delta_{ab}}{12}
\Bigg
[\Bigg(\langle\bar{q}q\rangle_{\rho_N}+x^{\mu}\langle\bar{q}D_{\mu}q\rangle_{
\rho_N}
 \nonumber \\
&+&\frac{1}{2}x^{\mu}x^{\nu}\langle\bar{q}D_{\mu}D_{\nu}q\rangle_{\rho_N}
+...\Bigg)\delta_{\alpha\beta}\nonumber \\
&&+\Bigg(\langle\bar{q}\gamma_{\lambda}q\rangle_{\rho_N}+x^{\mu}\langle\bar{q}
\gamma_{\lambda}D_{\mu} q\rangle_{\rho_N}
 \nonumber \\
&+&\frac{1}{2}x^{\mu}x^{\nu}\langle\bar{q}\gamma_{\lambda}D_{\mu}D_{\nu}
q\rangle_{\rho_N}
+...\Bigg)\gamma^{\lambda}_{\alpha\beta} \Bigg].\nonumber \\
\end{eqnarray}
The quark-gluon mixed condensate in nuclear matter is written as
\begin{eqnarray} \label{ }
&&\langle
g_{s}q_{a\alpha}\bar{q}_{b\beta}G_{\mu\nu}^{A}\rangle_{\rho_N}\nonumber \\&&=-\frac{t_{
ab}^{A }}{96}\Bigg\{\langle g_{s}\bar{q}\sigma\cdot
Gq\rangle_{\rho_N}
\Bigg[\sigma_{\mu\nu}+i(u_{\mu}\gamma_{\nu}-u_{\nu}\gamma_{\mu
})
\!\not\! {u}\Bigg]_{\alpha\beta} \nonumber \\
&&+\langle g_{s}\bar{q}\!\not\! {u}\sigma\cdot
Gq\rangle_{\rho_N}
 \Bigg[\sigma_{\mu\nu}\!\not\!
{u}+i(u_{\mu}\gamma_{\nu}-u_{\nu}\gamma_{\mu}
)\Bigg]_{\alpha\beta} \nonumber \\
&&-4\Bigg(\langle\bar{q}u\cdot D u\cdot D
q\rangle_{\rho_N}
+im_{q}\langle\bar{q}
\!\not\! {u}u\cdot D q\rangle_{\rho_N}\Bigg) \nonumber \\
&&\times\Bigg[\sigma_{\mu\nu}+2i(u_{\mu}\gamma_{\nu}-u_{\nu}\gamma_{\mu}
)\!\not\! {u}\Bigg]_{\alpha\beta}\Bigg\},
\nonumber \\
\end{eqnarray}
where $t_{ab}^{A}$ are  Gell-Mann matrices and
$D_\mu=\frac{1}{2}(\gamma_\mu \!\not\!{D}+\!\not\!{D}\gamma_\mu)$.
The matrix element of the four-dimension gluon condensate can also be
parameterized as 
\begin{eqnarray}
 \langle
G_{\kappa\lambda}^{A}G_{\mu\nu}^{B}\rangle_{\rho_N}&=&\frac{\delta^{AB}}{96}
\Bigg[ \langle
G^{2}\rangle_{\rho_N}(g_{\kappa\mu}g_{\lambda\nu}-g_{\kappa\nu}g_{\lambda\mu}
)
\nonumber \\
&+&O(\langle \textbf{E}^{2}+\textbf{B}^{2}\rangle_{\rho_N})\Bigg],
\end{eqnarray}
where we ignore from the last term in this equation because of its
negligible contribution. The different condensates in the above equations
are defined in the following way \cite{Cohen:1994wm,Jin:1992id}:
\begin{eqnarray} \label{ }
\langle\bar{q}\gamma_{\mu}q\rangle_{\rho_N}&=&\langle\bar{q}\!\not\!{u}q\rangle_
{\rho_N}
u_{\mu} ,\nonumber \\
\langle\bar{q}D_{\mu}q\rangle_{\rho_N}&=&\langle\bar{q}u\cdot D
q\rangle_{\rho_N}
u_{\mu}=-im_{q}\langle\bar{q}\!\not\!{u}q\rangle_{\rho_N}
u_{\mu}  ,\nonumber \\
\langle\bar{q}\gamma_{\mu}D_{\nu}q\rangle_{\rho_N}&=&\frac{4}{3}\langle\bar{q}
\!\not\! {u}u\cdot D
q\rangle_{\rho_N}(u_{\mu}u_{\nu}-\frac{1}{4}g_{\mu\nu})\nonumber \\
&+&\frac{i}{3}m_{q}
\langle\bar{q}q\rangle_{\rho_N}(u_{\mu}u_{\nu}-g_{\mu\nu}),
\nonumber \\
\langle\bar{q}D_{\mu}D_{\nu}q\rangle_{\rho_N}&=&\frac{4}{3}\langle\bar{q}
u\cdot D u\cdot D
q\rangle_{\rho_N}(u_{\mu}u_{\nu}-\frac{1}{4}g_{\mu\nu})\nonumber \\
&-&\frac{1}{6} \langle
g_{s}\bar{q}\sigma\cdot Gq\rangle_{\rho_N}(u_{\mu}u_{\nu}-g_{\mu\nu}) , \nonumber \\
\langle\bar{q}\gamma_{\lambda}D_{\mu}D_{\nu}q\rangle_{\rho_N}&=&2\langle\bar{q}
\!\not\! {u}u\cdot D u\cdot D q\rangle_{\rho_N}
\nonumber \\
&&\Bigg[u_{\lambda}u_{\mu}u_{\nu} -\frac{1}{6}
(u_{\lambda}g_{\mu\nu}+u_{\mu}g_{\lambda\nu}+u_{\nu}g_{\lambda\mu})\Bigg]\nonumber\\
&&-\frac{1}{6} \langle g_{s}\bar{q}\!\not\! {u}\sigma\cdot
Gq\rangle_{\rho_N}(u_{\lambda}u_{\mu}u_{\nu}-u_{\lambda}g_{\mu\nu}),\nonumber
\\
\end{eqnarray}
where, in their derivations,  the  equation of motion has been used and the terms
$\textit{O}(m^2_q)$  have been neglected due to
 their ignorable  contributions \cite{Cohen:1994wm}.

 By substituting the above matrix elements and the in-medium condensates, after lengthy calculations, we find the expression of the correlation function in
 coordinate space. Using the relation,
\begin{eqnarray} \label{x2n}
\frac{1}{(x^2)^n}&=&\int\frac{d^Dt}{(2\pi)^D}e^{-it\cdot
x}i(-1)^{n+1}2^{D-2n}\pi^{D/2} \nonumber
\\
&&\times\frac{\Gamma(D/2-n)}{\Gamma(n)}(-\frac{1}{t^2})^{D/2-n},
\end{eqnarray}
We transform the calculations to the momentum space. Then, by the help of the replacement
\begin{eqnarray} \label{Log}
\Gamma\Big(\frac{D}{2}-n\Big)\Big(-\frac{1}{L}\Big)^{\frac{D}{2}-n}\rightarrow
\frac{(-1)^{n-1}}{(n-2)!}(-L)^{n-2}\ln(-L),\nonumber
\\
\end{eqnarray}
we find  the imaginary parts of the obtained results for different structures called
the spectral densities $\Delta\rho_i(p_0,\vec{p})$ in OPE side in terms of
$(p^2)^n$. After ordering the Dirac matrices like the physical side, we set
$p^2=p_0^2-|\vec{p}|^2$ and replace $p_0$ with $w$. In order to
remove the contributions of the negative energy particles, we
multiply the OPE side by the weight function
$(w-\bar{E}_p)e^{-\frac{w^2}{M^2}}$ like the physical side and
perform the integral
\begin{eqnarray} \label{Pi1}
\Pi_i(w_0,\vec{p})=\int^{w_0}_{-w_0} dw
\Delta\rho_i(w,\vec{p})(w-\bar{E}_p)e^{-\frac{w^2}{M^2}}.
\end{eqnarray} \label{Pi1}
By carrying out the integration over $w$, one can find the
$\Pi_i(w_0,\vec{p})$ functions in Borel scheme. By using
$w_0=\sqrt{s_0^*}$, with $s_0^*$ being the continuum threshold in
nuclear matter, and making some variable changing, we find the
final expressions of the $\Pi_i(s_0^*,M^2)$ functions. As an
example, we present the functions $\Pi_{i}(s_0^*,M^2)$ for
$\Sigma^{*}$ which are obtained as
\begin{widetext}
\begin{eqnarray}\label{A1}
\Pi_{i}(s_0^*,M^2)=\Pi_{i}^{pert}(s_0^*,M^2)+\sum_{k=3}^{k=6}\Pi_{i}^{k}(s_0^*,M^2),
\end{eqnarray}
where $ ``pert" $ denotes the perturbative contributions and  the upper indices $ 3 $, $ 4 $, $ 5 $ and $ 6 $ stand for the nonperturbative contributions. These functions are obtained as
\begin{eqnarray}\label{A2}
\Pi_{1}^{pert}(s_0^*,M^2)&=&\frac{1}{512 \pi^{4}} \left[3\bar{E}_p
M^2\sqrt{s_0^*}(m_d+m_u+m_s)(3M^2-4\vec{p}^2+2s_0^{*})\right]
e^{-\frac{s_0^*}{M^2}}
\nonumber \\
&-&\frac{1}{1024 \pi^4}\int_{0}^{s_0^*}ds \frac{3 \bar{E}_p
(m_d+m_u+m_s)(3M^4-4M^2 \vec{p}^2+4\vec{p}^4)
}{\sqrt{s}}e^{-\frac{s}{M^2}},
\nonumber \\
\Pi_{2}^{pert}(s_0^*,M^2)&=&\frac{1}{640 \pi^{4}} \left[\bar{E}_p
M^2\sqrt{s_0^*}(3M^2-4\vec{p}^2+2s_0^*)\right]e^{-\frac{s_0^*}{M^2}}
\nonumber \\
&-&\frac{1}{1280 \pi^4}\int_{0}^{s_0^*}ds \frac{ \bar{E}_p
(3M^4-4M^2 \vec{p}^2+4\vec{p}^4) }{\sqrt{s}}e^{-\frac{s}{M^2}},
\nonumber \\
\Pi_{3}^{pert}(s_0^*,M^2)&=&0,
\end{eqnarray}
\begin{eqnarray}\label{A3}
\Pi_{1}^{3}(s_0^*,M^2)&=&\frac{M^2\sqrt{s_0^*}}{24\pi^2}\Big[\Big(3m_s+3m_d-4m_q\Big)
\langle u^{\dag}u\rangle_{\rho_N}+\Big(3m_u+3m_s-4m_q\Big)\langle
d^{\dag}d\rangle_{\rho_N}+\Big(3m_u+3m_d-4m_s\Big)\langle
s^{\dag}s\rangle_{\rho_N}
\nonumber \\
&-&2\bar{E}_p(\langle \bar{s}s\rangle_{\rho_N}+\langle
\bar{u}u\rangle_{\rho_N}+\langle
\bar{d}d\rangle_{\rho_N})\Big]e^{-\frac{s_0^*}{M^2}}
\nonumber \\
&+&\frac{1}{144
\pi^{2}}\int_{0}^{s_0^*}ds\frac{1}{\sqrt{s}}\Big[4\bar{E}_p\Big(\langle
\bar{d}iD_0iD_0d\rangle_{\rho_N}+\langle
\bar{u}iD_0iD_0u\rangle_{\rho_N}+\langle
\bar{s}iD_0iD_0s\rangle_{\rho_N}\Big)-4\bar{E}_p\Big(\langle
\bar{d}g_s\sigma G d\rangle_{\rho_N}
\nonumber \\
&+&\langle\bar{u}g_s\sigma G u\rangle_{\rho_N}+\langle
\bar{s}g_s\sigma G
s\rangle_{\rho_N}\Big)-12\bar{E}_p(m_u+m_s)\langle
d^{\dag}iD_0d\rangle_{\rho_N}-12\bar{E}_p(m_d+m_s)\langle
u^{\dag}iD_0u\rangle_{\rho_N}
\nonumber \\
&-& 12\bar{E}_p(m_d+m_u)\langle
s^{\dag}iD_0s\rangle_{\rho_N}-6\bar{E}_p(m_qm_s+m_qm_u-M^2+2\vec{p}^2)\langle
\bar{d}d\rangle_{\rho_N}
\nonumber \\
&-& 6\bar{E}_p(m_qm_s+m_qm_d-M^2+2\vec{p}^2)\langle
\bar{u}u\rangle_{\rho_N}-6\bar{E}_p(m_qm_u+m_qm_d-M^2+2\vec{p}^2)\langle
\bar{s}s\rangle_{\rho_N}
\nonumber \\
&+& (12m_q-9m_s-9m_u)\langle
d^{\dag}d\rangle_{\rho_N}+(12m_q-9m_s-9m_d)\langle
u^{\dag}u\rangle_{\rho_N}+(12m_q-9m_d-9m_u)\langle
s^{\dag}s\rangle_{\rho_N} \Big]e^{-\frac{s}{M^2}} ,
\nonumber \\
\Pi_{2}^{3}(s_0^*,M^2)&=&\frac{M^2\sqrt{s_0^*}}{36\pi^2}\Big(\langle
u^{\dag}u\rangle_{\rho_N}+\langle
d^{\dag}d\rangle_{\rho_N}+\langle s^{\dag}s\rangle_{\rho_N}\Big)
\nonumber \\
&+&\frac{1}{216
\pi^{2}}\int_{0}^{s_0^*}ds\frac{1}{\sqrt{s}}\Big[4\bar{E}_p\Big(\langle
d^{\dag}iD_0d\rangle_{\rho_N}+\langle
u^{\dag}iD_0u\rangle_{\rho_N}+\langle
s^{\dag}iD_0s\rangle_{\rho_N}\Big)+\bar{E}_p(27m_s+27m_u-10m_q)\langle
\bar{d}d\rangle_{\rho_N}
\nonumber \\
&+&\bar{E}_p(27m_s+27m_d-10m_q)\langle
\bar{u}u\rangle_{\rho_N}+\bar{E}_p(27m_u+27m_d-10m_q)\langle
\bar{s}s\rangle_{\rho_N}
\nonumber \\
&-&3M^2\Big(\langle u^{\dag}u\rangle_{\rho_N}+\langle
d^{\dag}d\rangle_{\rho_N}\langle+
s^{\dag}s\rangle_{\rho_N}\Big)\Big]e^{-\frac{s}{M^2}} ,
\nonumber \\
\Pi_{3}^{3}(s_0^*,M^2)&=&\frac{M^2\sqrt{s_0^*}}{216
\pi^2}\Big[-32\Big(\langle u^{\dag}iD_0u\rangle_{\rho_N}+\langle
d^{\dag}iD_0d\rangle_{\rho_N} +\langle
s^{\dag}iD_0s\rangle_{\rho_N}\Big)-9 \bar{E}_p\Big(\langle
u^{\dag}u\rangle_{\rho_N} +\langle
d^{\dag}d\rangle_{\rho_N}+\langle s^{\dag}s\rangle_{\rho_N}\Big)
\nonumber \\
&+&8m_q\Big(\langle \bar{u}u \rangle_{\rho_N}+\langle \bar{d}d
\rangle_{\rho_N}+\langle \bar{s}s \rangle_{\rho_N}\Big)\Big]
\nonumber \\
&+&\frac{1}{432 \pi^2}\int_{0}^{s_0^*}ds
\frac{1}{\sqrt{s}}\Big[12\bar{E}_p\Big(\langle
d^{\dag}iD_0iD_0d\rangle_{\rho_N}+\langle
u^{\dag}iD_0iD_0u\rangle_{\rho_N}+\langle
s^{\dag}iD_0iD_0s\rangle_{\rho_N}\Big)
\nonumber \\
&+&32M^2\Big(\langle d^{\dag}iD_0d\rangle_{\rho_N}+\langle
u^{\dag}iD_0u\rangle_{\rho_N}+\langle
s^{\dag}iD_0s\rangle_{\rho_N}\Big)-8M^2mq\Big(\langle
\bar{d}d\rangle_{\rho_N}+\langle \bar{u}u\rangle_{\rho_N}+\langle
\bar{s}s\rangle_{\rho_N}\Big)
\nonumber \\
&-& 7\bar{E}_p\Big(\langle d^{\dag}g_s\sigma
Gd\rangle_{\rho_N}+\langle u^{\dag}g_s\sigma
Gu\rangle_{\rho_N}+\langle s^{\dag}g_s\sigma
Gs\rangle_{\rho_N}\Big) -\bar{E}_p
(54m_qm_s+54m_qm_u-9M^2+18\vec{p}^2)\langle
d^{\dag}d\rangle_{\rho_N}
\nonumber \\
&-&\bar{E}_p (54m_qm_d+54m_qm_s-9M^2+18\vec{p}^2)\langle
u^{\dag}u\rangle_{\rho_N}-\bar{E}_p
(54m_qm_d+54m_qm_u-9M^2+18\vec{p}^2)\langle
s^{\dag}s\rangle_{\rho_N}\Big]e^{-\frac{s}{M^2}} \nonumber \\,
\end{eqnarray}
\begin{eqnarray}\label{A4}
\Pi_{1}^{4}(s_0^*,M^2)&=&\frac{1}{128 \pi^{2}}
\langle\frac{\alpha_s G^2}{\pi}
\rangle_{\rho_N}\int_{0}^{s_0^*}dw\frac{ \bar{E}_p (m_d+m_u+m_s)
}{\sqrt{w}}e^{-\frac{s}{M^2}},
\nonumber \\
\Pi_{2}^{4}(s_0^*,M^2)&=&\frac{1}{576 \pi^{2}}
\langle\frac{\alpha_s G^2}{\pi}
\rangle_{\rho_N}\int_{0}^{s_0^*}ds\frac{ \bar{E}_p
 }{\sqrt{s}}e^{-\frac{s}{M^2}},
 \nonumber \\
\Pi_{3}^{4}(s_0^*,M^2)&=&0,
\end{eqnarray}
\begin{eqnarray}\label{A5}
\Pi_{1}^{5}(s_0^*,M^2)&=&\frac{1}{48 \pi^{2}} \Big[4m_q\langle
s^{\dag}iD_0 s \rangle_{\rho_N}+4m_q \langle d^{\dag}iD_0 d
\rangle_{\rho_N}+4m_q\langle u^{\dag}iD_0 u
\rangle_{\rho_N}-4\langle \bar{d}iD_0iD_0d
\rangle_{\rho_N}-4\langle \bar{s}iD_0iD_0s \rangle_{\rho_N}
 \nonumber \\
 &-&\langle \bar{u}iD_0iD_0u
\rangle_{\rho_N}-\langle \bar{d}g_s\sigma G d
\rangle_{\rho_N}-\langle \bar{s}g_s\sigma G s
\rangle_{\rho_N}-\langle \bar{u}g_s\sigma G u
\rangle_{\rho_N}\Big]\int_{0}^{s_0^*}ds\frac{ \bar{E}_p
 }{\sqrt{s}}e^{-\frac{s}{M^2}},
 \nonumber \\
\Pi_{2}^{5}(s_0^*,M^2)&=&0,
\nonumber \\
\Pi_{3}^{5}(s_0^*,M^2)&=&-\frac{1}{72 \pi^{2}} \Big[\langle
d^{\dag}g_s\sigma G d \rangle_{\rho_N}+\langle u^{\dag}g_s\sigma G
u \rangle_{\rho_N}+\langle s^{\dag}g_s\sigma G s
\rangle_{\rho_N}\Big]\int_{0}^{s_0^*}ds\frac{ \bar{E}_p
 }{\sqrt{s}}e^{-\frac{s}{M^2}},
 \end{eqnarray}
\begin{eqnarray}\label{A6}
\Pi_{1}^{6}(s_0^*,M^2)&=&0,
\nonumber \\
\Pi_{2}^{6}(s_0^*,M^2)&=&0,
\nonumber \\
\Pi_{3}^{6}(s_0^*,M^2)&=&0.
\end{eqnarray}

\end{widetext}

\subsection{Sum Rules for Physical Observables: Numerical Results }

Having obtained the hadronic and OPE sides of the correlation
function, we match them to find QCD sum rules for the mass, residue
and self energies of the considered decuplet baryons:
\begin{eqnarray}\label{HadOPE}
\lambda_D^{*2}m_D^* e^{-\frac{E_p^2}{M^2}}&=&\Pi_{1}(s_0^*,M^2),
\nonumber \\
\lambda_D^{*2} e^{-\frac{E_p^2}{M^2}}&=&\Pi_{2}(s_0^*,M^2),
\nonumber \\
\lambda_D^{*2}\Sigma_{\nu}
e^{-\frac{E_p^2}{M^2}}&=&\Pi_{3}(s_0^*,M^2).
\end{eqnarray}

Now, we proceed to numerically analyze the above sum rules in  $\Delta^0, \Sigma^{*},
\Xi^{*}$ and $\Omega^{-}$ channels both in vacuum and nuclear medium. The
 sum rules contain numerous parameters, numerical values of which are collected  in table \ref{inputpar}.
\begin{widetext}
\begin{table}[ht!]
\centering
%\rowcolors{1}{lightgray}{white}
\begin{tabular}{ll}
\hline \hline
   Input parameters  &  Values
           \\
\hline \hline
$\mid\vec{p}\mid   $          &  $270~MeV $    \cite{Cohen:1994wm}  \\
$ m_{u}   $ ; $ m_{d}   $ ;  $ m_{s}   $  &  $2.2_{-0.4}^{0.6}  $ $MeV$
; $4.7_{-0.4}^{0.5}  $ $MeV$ ;   $96^{+8}_{-4}  $ $MeV$   \cite{PDG}      \\
$ \rho_{N}     $          &  $(0.11)^3  $ $GeV^3$    \cite{Cohen:1994wm,Jin:1992id,Cohen:1991nk}     \\
$ \langle q^{\dag}q\rangle_{\rho_N}    $ ; $ \langle
s^{\dag}s\rangle_{\rho_N}    $
   &  $\frac{3}{2}\rho_{N}$   ; 0    \cite{Cohen:1994wm,Jin:1992id,Cohen:1991nk,Wang:2012xk}   \\
$ \langle\bar{q}q\rangle_{0} $  ; $ \langle\bar{s}s\rangle_{0} $ &
$ (-0.241)^3
 $ $GeV^3$ ; 0.8 $ \langle\bar{q}q\rangle_{0} $   \cite{Belyaev}        \\
$ m_{q}      $          &  $0.5(m_{u}+m_{d})$           \cite{Cohen:1994wm,Jin:1992id,Cohen:1991nk}      \\
$ \sigma_{N} $      &  $0.059 ~  $GeV$ $
 \cite{Alarcon:2011zs}  \\
$y$ & $0.04\pm0.02$ \cite{Thomas:2012tg}; $0.066\pm0.011\pm0.002$ \cite{Dinter:2011za};
$ 0.02(13)(10) $ \cite{Alarcon:2012nr}\\
$  \langle\bar{q}q\rangle_{\rho_N}  $ ;  $
\langle\bar{s}s\rangle_{\rho_N}  $  &  $
\langle\bar{q}q\rangle_{0}+\frac{\sigma_{N}}{2m_{q}}\rho_{N}$
; $ \langle\bar{s}s\rangle_{0}+y\frac{\sigma_{N}}{2m_{q}}\rho_{N}$    \cite{Cohen:1994wm,Jin:1992id,Cohen:1991nk,Jin:1993up,Wang:2012xk}              \\
$  \langle q^{\dag}g_{s}\sigma Gq\rangle_{\rho_N}  $ ;  $  \langle
s^{\dag}g_{s}\sigma Gs\rangle_{\rho_N} $ &
  $ -0.33~GeV^2 \rho_{N}$ ; $ -y 0.33~GeV^2 \rho_{N}$ \cite{Cohen:1994wm,Jin:1992id,Cohen:1991nk,Jin:1993up,Wang:2012xk}\\
$  \langle q^{\dag}iD_{0}q\rangle_{\rho_N}  $    ; $  \langle
s^{\dag}iD_{0}s\rangle_{\rho_N}  $      &  $0.18 ~GeV \rho_{N}$ ;$
\frac{m_s \langle\bar{s}s\rangle_{\rho_N}}{4}
+0.02 ~GeV \rho_N$            \cite{Cohen:1994wm,Jin:1992id,Cohen:1991nk,Jin:1993up,Wang:2012xk}    \\
$  \langle\bar{q}iD_{0}q\rangle_{\rho_N}  $    ; $
\langle\bar{s}iD_{0}s\rangle_{\rho_N}  $      &  $\frac{3}{2} m_q
\rho_{N}\simeq0 $ ; 0
\cite{Cohen:1994wm,Jin:1992id,Cohen:1991nk,Jin:1993up,Wang:2012xk}\\
$  m_{0}^{2}  $          &  $ 0.8~GeV^2$    \cite{Belyaev}               \\
$   \langle\bar{q}g_{s}\sigma Gq\rangle_{0} $ ; $   \langle\bar{s}g_{s}\sigma Gs\rangle_{0} $  &  $m_{0}^{2}\langle\bar{q}q\rangle_{0} $ ; $m_{0}^{2}\langle\bar{s}s\rangle_{0} $ \\
$  \langle\bar{q}g_{s}\sigma Gq\rangle_{\rho_N}  $  ;$
\langle\bar{s}g_{s}\sigma Gs\rangle_{\rho_N}  $ &
$\langle\bar{q}g_{s}\sigma Gq\rangle_{0}+3~GeV^2\rho_{N} $ ;
$\langle\bar{s}g_{s}\sigma Gs\rangle_{0}+3 y ~GeV^2\rho_{N} $\cite{Cohen:1994wm,Jin:1992id,Cohen:1991nk,Jin:1993up,Wang:2012xk} \\
$ \langle  \bar{q}iD_{0}iD_{0}q\rangle_{\rho_N} $  ;$\langle \bar{s}iD_{0}iD_{0}s\rangle_{\rho_N} $  &  $ 0.3~GeV^2\rho_{N}-\frac{1}{8}\langle\bar{q}g_{s}\sigma Gq\rangle_{\rho_N}$ ; \\
& $ 0.3 y ~GeV^2\rho_{N}-\frac{1}{8}\langle\bar{s}g_{s}\sigma Gs\rangle_{\rho_N}$  \cite{Cohen:1994wm,Jin:1992id,Cohen:1991nk,Jin:1993up,Wang:2012xk}\\
$  \langle q^{\dag}iD_{0}iD_{0}q\rangle_{\rho_N}  $ ;$  \langle s^{\dag}iD_{0}iD_{0}s\rangle_{\rho_N}$ & $0.031~GeV^2\rho_{N}-\frac{1}{12}\langle q^{\dag}g_{s}\sigma Gq\rangle_{\rho_N} $; \\
& $0.031 y~GeV^2\rho_{N}-\frac{1}{12}\langle s^{\dag}g_{s}\sigma Gs\rangle_{\rho_N} $ \cite{Cohen:1994wm,Jin:1992id,Cohen:1991nk,Jin:1993up,Wang:2012xk}\\
$\langle \frac{\alpha_s}{\pi} G^{2}\rangle_{0}$ & $(0.33\pm0.04)^4~GeV^4$ \cite{Belyaev}\\
$\langle \frac{\alpha_s}{\pi} G^{2}\rangle_{\rho_N}$ & $\langle \frac{\alpha_s}{\pi} G^{2}\rangle_{0}-0.65~GeV \rho_N$ \cite{Cohen:1994wm,Jin:1992id,Cohen:1991nk}\\
 \hline \hline
\end{tabular}
\caption{Numerical values of input parameters.}\label{inputpar}
\end{table}

\end{widetext}
Besides the above input parameters, the QCD sum rules depend also
on two auxiliary parameters that should be fixed: the Borel
parameter $M^2$ and the continuum threshold $s^*_0$.  The
continuum threshold   is not totally arbitrary and it is correlated with
 the energy of the first excited state with the same quantum
numbers as the  interpolating currents for decuplet baryons. According to the standard prescriptions, we take the interval $(m_D+0.4)^2~GeV^2\leq
s^*_0\leq(m_D+0.6)^2~GeV^2$. The standard criteria in calculating
the working window of the Borel parameter is that not only the contributions of the higher
resonances and continuum should be adequately suppressed, but the contributions of the higher
dimensional condensates should be small and the perturbative contributions should exceed the nonperturbative ones. These criteria lead to the following intervals:
\begin{eqnarray*}\label{MsqRange}
1.1~GeV^2\leq&M^2&\leq1.4~GeV^2~~~~~~~~\mathrm{for} ~~\Delta^0
\nonumber \\
1.5~GeV^2\leq&M^2&\leq1.9~GeV^2~~~~~~~~\mathrm{for} ~~\Sigma^{^*0}
\nonumber \\
2.2~GeV^2\leq&M^2&\leq2.5~GeV^2~~~~~~~~\mathrm{for} ~~\Xi^{*}
\nonumber \\
2.6~GeV^2\leq&M^2&\leq3.0~GeV^2~~~~~~~~\mathrm{for} ~~\Omega^{-}
\end{eqnarray*}

Making use of the  working windows of the auxiliary parameters and the values of
other inputs,  as examples, we plot the in-medium mass, $ m^{*}_{\Delta} $, residue, $ \lambda^{*}_{\Delta} $, and  vector self energy, $\Sigma^{\nu}_{ \Delta}$, of the  $\Delta$
baryon as   functions of  $%
M^2$ at different fixed values  of the threshold parameter $s_0$ and central values of other input parameters in figures 1-3. From these figures we see that the in-medium mass and residue as well as the vector self energy demonstrate good  stability with respect to  $%
M^2$ in its working region. It is also clear that the results very weakly depend on the threshold parameter $s_0$ in its working window. 
\begin{figure}
\includegraphics[width=7.cm]{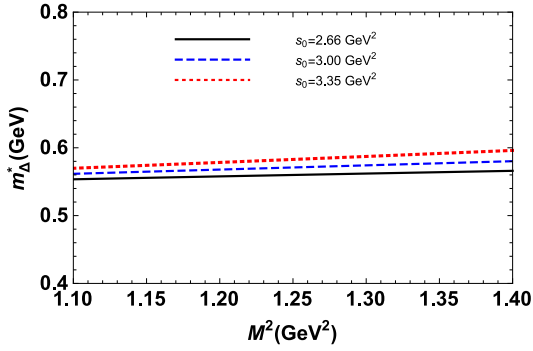}
\caption{The in-medium mass of the $\Delta$
baryon as  a function of  $%
M^2$ at different fixed values  of the threshold parameter $s_0$ and central values of other input parameters.}
\label{fig:SC2}
\end{figure}
\begin{figure}
\includegraphics[width=7.cm]{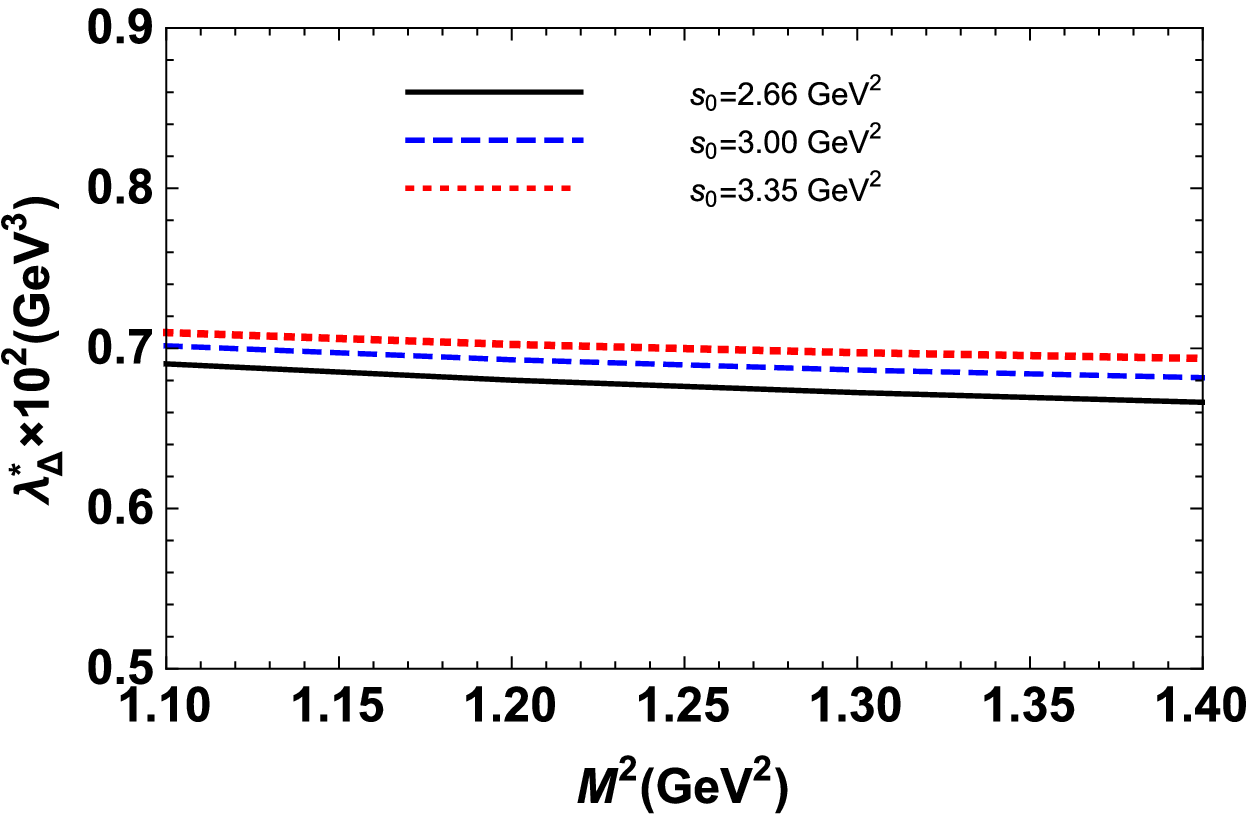}
\caption{The in-medium residue of the $\Delta$
baryon as  a function of  $%
M^2$ at different fixed values  of the threshold parameter $s_0$ and central values of other input parameters.}
\label{fig:SC2}
\end{figure}
\begin{figure}
\includegraphics[width=7.cm]{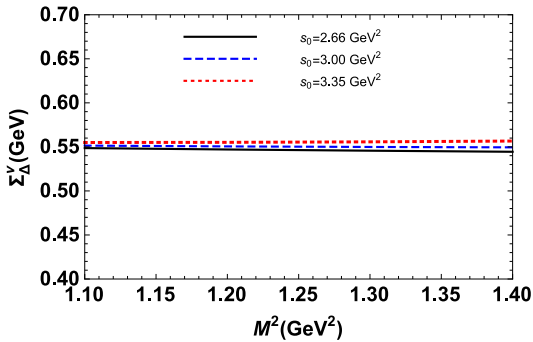}
\caption{The vector self energy of the $\Delta$
baryon as  a function of  $%
M^2$ at different fixed values  of the threshold parameter $s_0$ and central values of other input parameters.}
\label{fig:SC2}
\end{figure}

\begin{figure}
\includegraphics[width=7.cm]{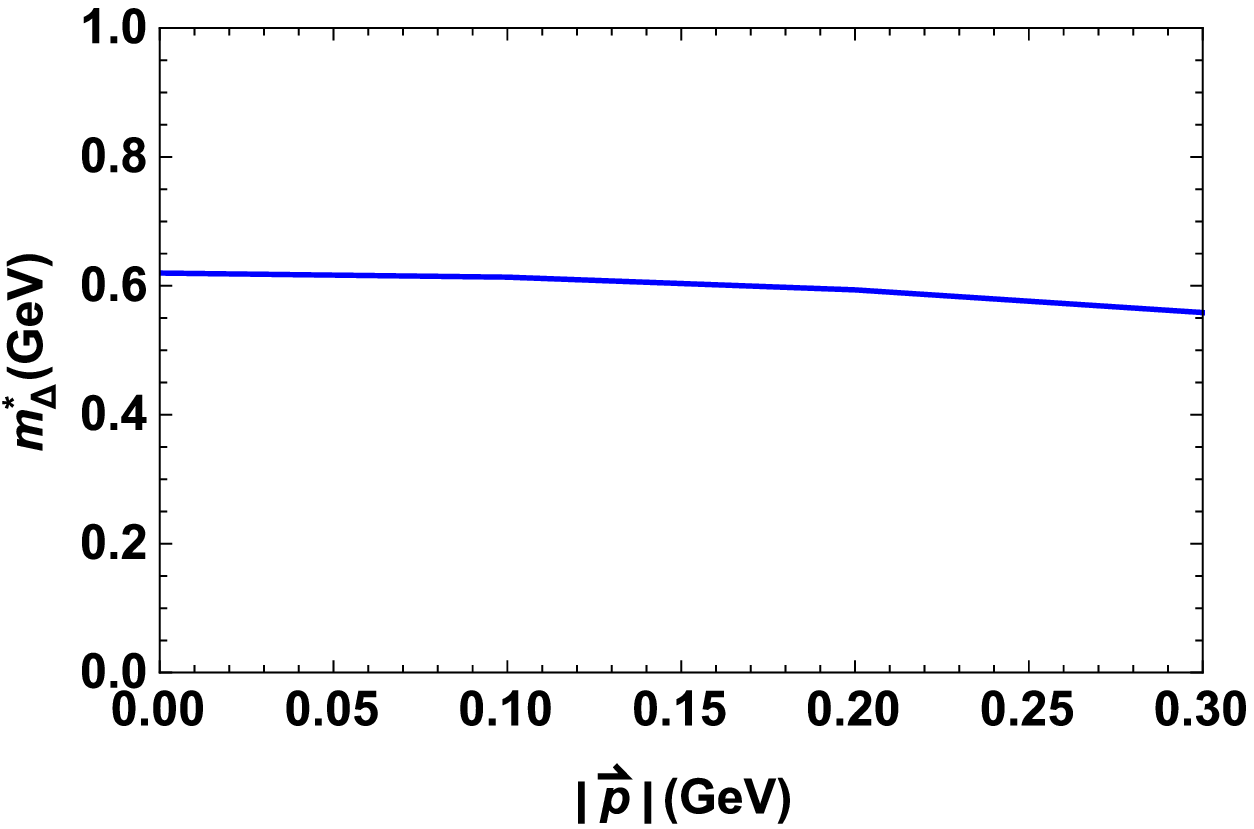}
\caption{The in-medium mass of the $\Delta$
baryon as  a function of  $\mid\vec{p}\mid$ at central  values of all auxiliary and input parameters.}
\label{fig:SC2}
\end{figure}
\begin{figure}
\includegraphics[width=7.cm]{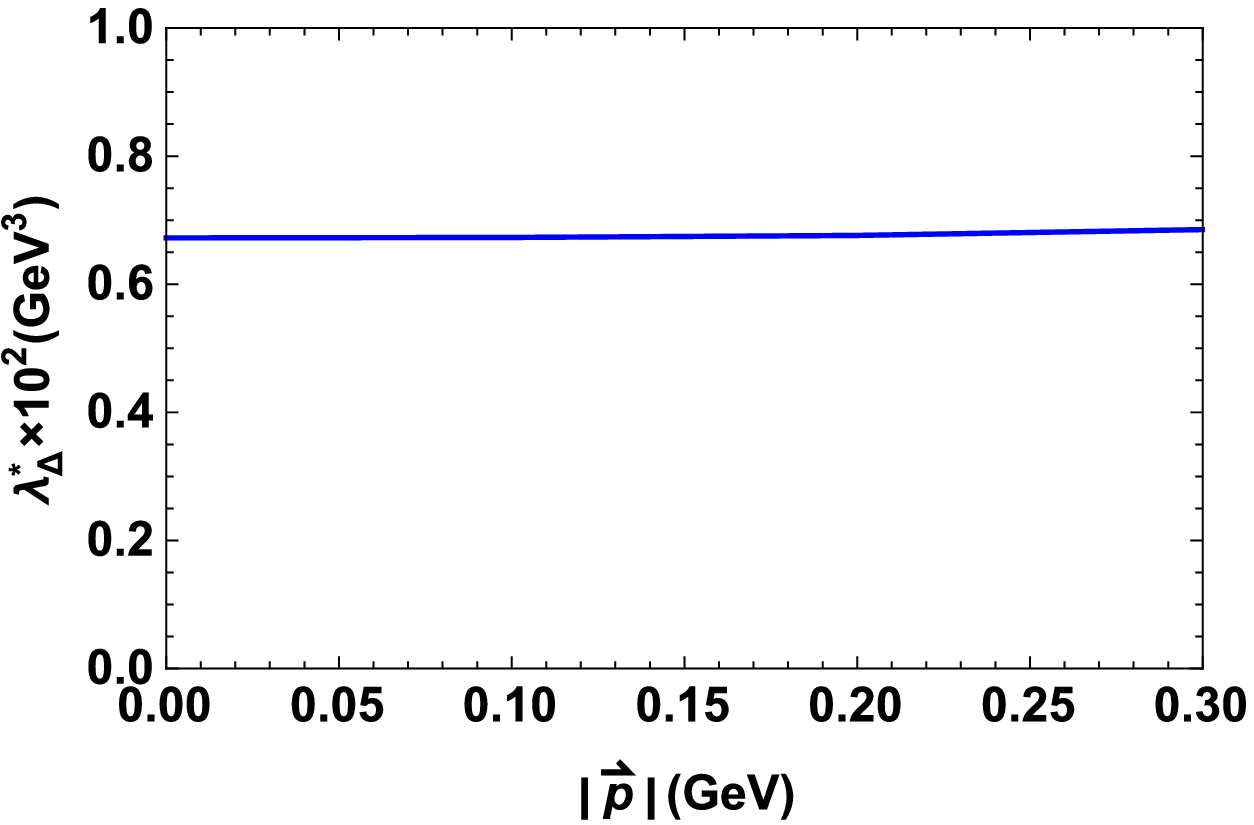}
\caption{The in-medium residue of the $\Delta$
baryon as  a function of  $\mid\vec{p}\mid$ at central  values of all auxiliary and input parameters.}
\label{fig:SC2}
\end{figure}
\begin{figure}
\includegraphics[width=7.cm]{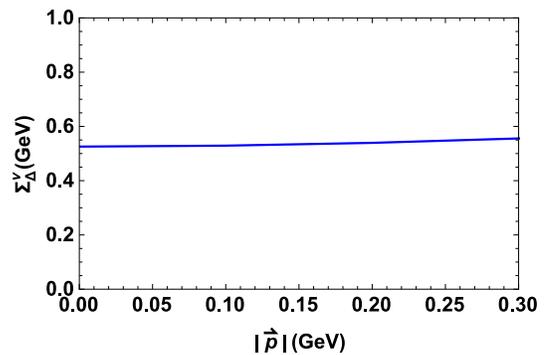}
\caption{The vector self-energy of the $\Delta$
baryon as  a function of  $\mid\vec{p}\mid$ at central  values of all auxiliary and input parameters.}
\label{fig:SC2}
\end{figure}

In this part, we would like to briefly discuss the dependence of the results on the values of the three-momentum of the particles under consideration and the density of the nuclear matter. We work at zero temperature and, as is seen from table  \ref{inputpar}, we take the external three-momentum of the quasi-particles approximately equal to Fermi momentum,   $\mid\vec{p}\mid = 270~MeV $, in the numerical analysis.  However, our numerical results show that the physical quantities overall do not considerably depend on this parameter in the interval [0, 0.27] $ MeV $ (see figures 4-6). This is an expected result. In the case of nucleons in nuclear matter, each quasi-nucleon 
has its own quasi-Fermi sea, hence, the external three-momentum of the quasi-nucleon is set at Fermi momentum at $ \rho_{N} =0.16~ fm^{-3} =(110 ~MeV)^{3}$ \cite{Cohen:1994wm,Azizi:2014yea}. For a similar reason, the external three-momentum for the quasi-decuplet baryons, especially   the strange members,  can be easily set to zero. To see how the results behave with respect to the nuclear matter density,  we show the dependence of the ratio of the mass and residue of, for instance, the $ \Delta $ baryon   in nuclear matter ($ m^{*}_{\Delta} $, $ \lambda^{*}_{\Delta} $) to the mass and residue in vacuum  ($ m_{\Delta} $, $ \lambda_{\Delta} $) as well as $  \Sigma^{\nu}_{ \Delta}/m^{*}_{\Delta} $ on $ \rho_N/\rho^{sat}_N $, with 
$ \rho^{sat}_N= (0.11)^3 ~ GeV^3$ being the saturation density used in the analysis, in figures (7-9). From these figures we see that the results depend linearly on the nuclear matter density. 

\begin{figure}
\includegraphics[width=7.cm]{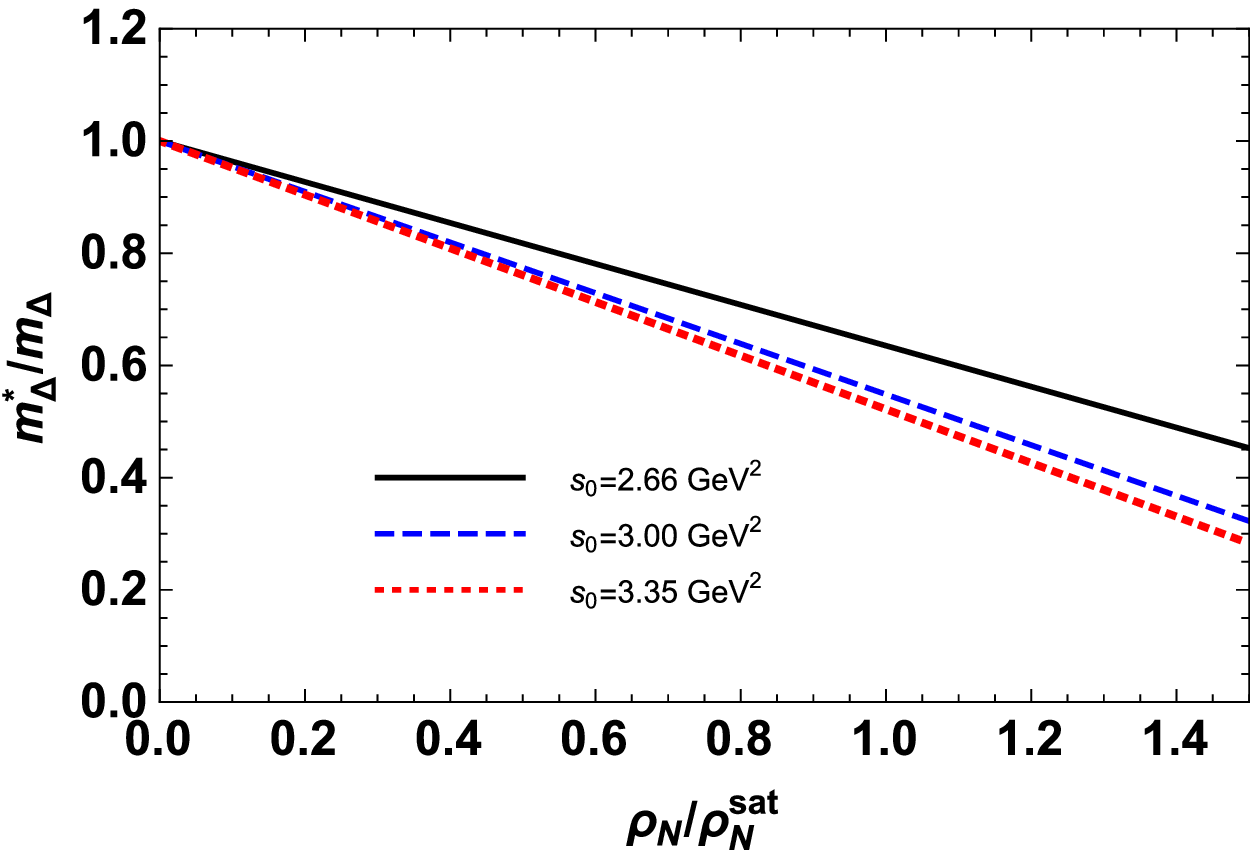}
\caption{$   m^{*}_{\Delta} /m_{\Delta} $ versus $ \rho_N/\rho^{sat}_N $ at central  values of $M^2$ and other input parameters.}
\label{fig:SC2}
\end{figure}
\begin{figure}
\includegraphics[width=7.cm]{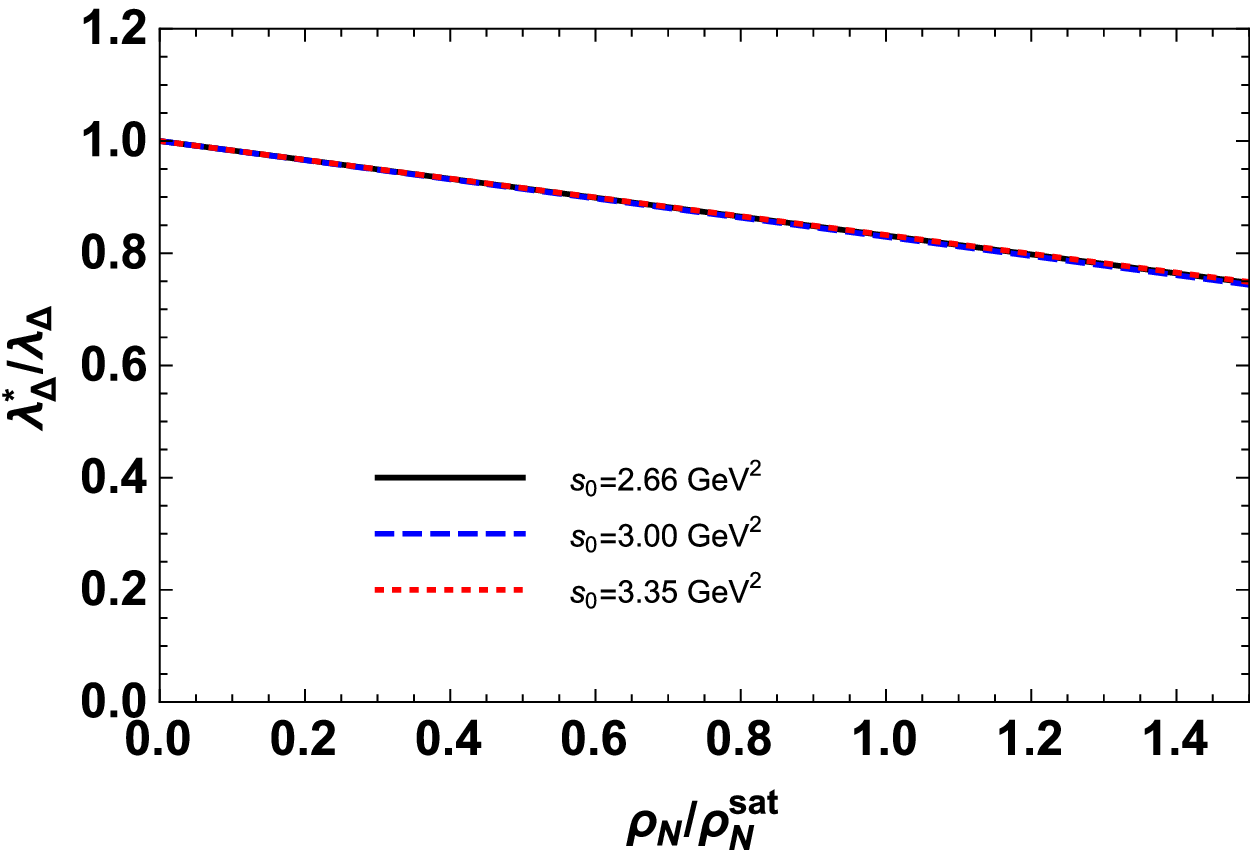}
\caption{$ \lambda^{*}_{\Delta}/ \lambda_{\Delta} $ versus $ \rho_N/\rho^{sat}_N $ at central  values of $M^2$ and other input parameters.}
\label{fig:SC2}
\end{figure}
\begin{figure}
\includegraphics[width=7.cm]{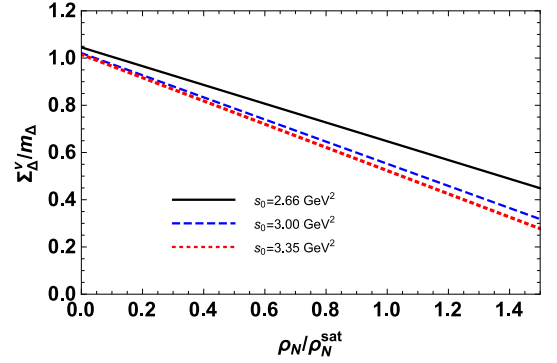}
\caption{$  \Sigma^{\nu}_{ \Delta}/m_{\Delta} $ versus $ \rho_N/\rho^{sat}_N $ at central  values of $M^2$ and other input parameters.}
\label{fig:SC2}
\end{figure}

After numerical analyses of the results for all baryons, using the values presented in table  \ref{inputpar}, we find the  values of the masses and  residues both in nuclear matter  and vacuum. We also obtain the vector   and scalar  self energies of the baryons under consideration in nuclear medium. Note that the vacuum results are obtained from those of the in-medium when $ \rho_N \rightarrow 0 $. The average values for the considered physical quantities   are presented in
table \ref{numericvalues}. The errors quoted in this table correspond to the uncertainties in the calculations of the working regions for the auxiliary parameters as well as those coming from the errors of other input parameters.

\begin{widetext}

\begin{table}[ht!]
\centering
%\rowcolors{1}{lightgray}{white}
\begin{tabular}{|c|c|c|c|c|c|c|}
\hline \hline
 &$\lambda_{\Delta}$ [GeV$^3]$  &$\lambda^{*}_{\Delta}$ [GeV$^3]$  & $m_{\Delta}$  [GeV]
  & $m^{*}_{\Delta}$  [GeV] & $\Sigma^{\nu}_{ \Delta}$ [MeV] & $\Sigma^{S}_{\Delta}$ [MeV] \\
 \hline
Present study &$ 0.013\pm0.004 $ &$ 0.007\pm 0.002$ &$ 1.297 \pm
0.364 $ & $ 0.571 \pm 0.159 $ & $  550\pm  51$ & -726    \\
\hline\hline
 &$\lambda_{\Sigma^{*}}$ [GeV$^3]$  &$\lambda^{*}_{\Sigma^{*}}$ [GeV$^3]$  & $m_{\Sigma^{*}}$  [GeV]
 & $m^{*}_{\Sigma^{*}}$  [GeV] & $\Sigma^{\nu}_{ \Sigma^{*}}$ [MeV] & $\Sigma^{S}_{\Sigma^{*}}$ [MeV] \\
 \hline
Present study & $ 0.024\pm0.007 $ & $ 0.016\pm 0.005$ & $ 1.385
\pm 0.387 $ & $ 0.927 \pm 0.259 $ &
 $  409\pm  41$ & -458
 \\  \hline\hline
 &$\lambda_{\Xi^{*}}$ [GeV$^3]$  &$\lambda^{*}_{\Xi^{*}}$ [GeV$^3]$  & $m_{\Xi^{*}}$  [GeV]
 & $m^{*}_{\Xi^{*}}$  [GeV] & $\Sigma^{\nu}_{ \Xi^{*}}$ [MeV] & $\Sigma^{S}_{\Xi^{*}}$ [MeV] \\
 \hline
Present study & $ 0.035\pm0.011 $ & $ 0.027\pm 0.008$ & $ 1.523
\pm 0.426 $ & $ 1.399 \pm 0.392 $ &
 $  148\pm  15$ & -124  \\
 \hline\hline
 &$\lambda_{\Omega^{-}}$ [GeV$^3]$  &$\lambda^{*}_{\Omega^{-}}$ [GeV$^3]$  & $m_{\Omega^{-}}$  [GeV]
 & $m^{*}_{\Omega^{-}}$  [GeV] & $\Sigma^{\nu}_{ \Omega^{-}}$ [MeV] & $\Sigma^{S}_{\Omega^{-}}$ [MeV] \\
 \hline
Present study & $ 0.044\pm0.013 $ & $ 0.042\pm 0.013$ & $ 1.668
\pm 0.467 $ & $ 1.634 \pm 0.457 $ &
 $  46\pm  5$ & -34  \\
 \hline \hline
\end{tabular}
\caption{The numerical values of masses, residues and
self-energies of  $\Delta$, $\Sigma^{*}$, $\Xi^{*}$ and
$\Omega^{-}$ baryons.}\label{numericvalues}
\end{table}

\end{widetext}

From this table, first of all, we see that our predictions on the masses in vacuum  are in
good consistencies with the average  experimental data presented in PDG \cite{PDG}. The masses obtained in the nuclear medium show negative shifts for all decuplet baryons. From the values of the scalar self energy ($ \Sigma^{S}_{D} $),  demonstrating the shifts in the masses due to finite density, we deduce that  the maximum shift in the masses, due to the nuclear medium, with amount of $56\% $ belongs to the $\Delta$ baryon and its minimum, $2\% $, corresponds to the  $\Omega^{-}$ state. This is an expected result since the $\Delta$ state have the same quark content as the nuclear medium and is more affected by the nuclear matter. When going from   
$\Delta$ to $\Omega^{-}$ the up and down quarks are replaced with the strange quark. The $\Omega^{-}$ state, having three $ s $ quarks, is less affected by the medium. The small shifts in the parameters of 
$\Omega^{-}$ may be attributed to the intrinsic strangeness in the nucleons.

 In the
case of the residues, our predictions  in vacuum are overall comparable with those obtained
in \cite{Lee:1997ix,Azizi:2016ddw} within the errors. The small differences may be linked to different input parameters used in these works. The values of residues  are also considerably affected by the medium. The shift in the residue of $\Delta$ with amount of $46\% $ is maximum. The residue of 
$\Omega^{-}$ again is minimally affected by the medium with amount of roughly $5\% $.

The value of vector self energy is considerably large in all decuplet channels.  It is again systematically  reduced when going  from the $\Delta$  to $\Omega^{-}$ baryon. Our results may be confronted with the experimental data of $\bar{P}$ANDA Collaboration at FAIR and NICA facility. However, we should remark that those experiments correspond to heavy ion collisions and  not exactly to a nuclear medium. Hence, the appropriate way to make such comparison would be to present sum rules at finite density but where the density is introduced through the baryonic chemical potential. This offers the possibility of exploring a wide range of densities. We worked with the nuclear matter density since the the in-medium condensates are available as functions of nuclear matter density not chemical potential and we extracted the zero-density (vacuum) sum rules, as a means of normalizing the finite density sum rules, to compare the results with the available experimental data and other theoretical predictions in vacuum.

\section{ACKNOWLEDGEMENTS}
K.~A. thanks  Do\v{g}u\c{s} University for the financial support through  the grant BAP
2015-16-D1-B04.

%%%%%%%%%%%%%%%%%%%%%%%%%%%%%%%%%%%%%%%%%%%%%%%%%%%%%%%%%%%%%%%%%%%%%%%%%%%%

\end{document}